\documentstyle[prd,aps,twocolumn]{revtex}
\input{psfig}
\begin{document}
\draft
\fnsymbol{footnote}

\wideabs{

\title{Relativistic Stellar Pulsations 
\\With Near-Zone Boundary Conditions\footnotemark
}

\author{Lee Lindblom}
\address{Theoretical Astrophysics 130-33,
         California Institute of Technology,
         Pasadena, CA 91125}

\author{Gregory Mendell and James R. Ipser}
\address{Department of Physics,
         University of Florida,
         Gainesville, FL 32611}

\date{\today}
\maketitle

\begin{abstract} A new method is presented here for evaluating
approximately the pulsation modes of relativistic stellar models. This
approximation relies on the fact that gravitational radiation influences
these modes only on timescales that are much longer than the basic
hydrodynamic timescale of the system.  This makes it possible to impose
the boundary conditions on the gravitational potentials at the surface
of the star rather than in the asymptotic wave zone of the gravitational
field.  This approximation is tested here by predicting the frequencies
of the outgoing non-radial hydrodynamic modes of non-rotating stars. 
The real parts of the frequencies are determined with an accuracy that
is better than our knowledge of the exact frequencies (about $0.01\%$)
except in the most relativistic models where it decreases to about
$0.1\%$.  The imaginary parts of the frequencies are determined with an 
accuracy of approximately $M/R$, where $M$ is the mass and $R$ is the
radius of the star in question. \end{abstract}

}

\narrowtext

\section{Introduction} 
\label{I}

The pulsation of stars in general relativity theory has been the subject
of scholarly investigation for many years.  Consequently there are
fairly well developed theories for modeling these oscillations
\cite{Thorne} \cite{LindDet} \cite{ChandraFerrari} \cite{Kokkotas}.  In
comparison with their Newtonian counterparts, these stellar pulsations
are fundamentally different, more interesting, and rather more difficult
to evaluate.  These differences arise because these pulsations couple to
gravitational radiation. The illumination of a star by gravitational
radiation, for example, causes that star to oscillate at the frequency
of the incoming radiation.  Thus, stars in general relativity theory can
oscillate at any frequency at all!

\footnotetext[1]{Published in Phys. Rev. D {\bf 56}, 2118 (1997).}

The {\it outgoing} (or more commonly but less descriptively, the
quasi-normal) modes of relativistic stars are defined by an additional
boundary condition: that they oscillate without incoming gravitational
radiation.  This boundary condition is simple enough to understand;
however, in practice it is rather difficult to impose numerically when
finding the modes of realistic relativistic stellar models.  This
difficulty arises because the external gravitational perturbations
must be decomposed into incoming and outgoing parts so that solutions
having only outgoing radiation can be identified.  This can only be
done simply in the region far away from the star in the wave zone of
the gravitational field.  In non-rotating stars it is straightforward
to integrate the exterior gravitational perturbations out into the
wave zone and perform this decomposition \cite{LindDet}.  In that case
the exterior gravitational field is the simple analytic Schwarzschild
geometry, and the gravitational perturbations obey reasonably simple
ordinary differential equations.  In the case of rotating stars,
however, this straightforward approach is not possible.  The exterior
geometries of rotating stars in general relativity are not known
analytically.  They are determined on numerical grids that become
asymptotically poor as they approach the wave zone.  The system of
partial differential equations that determines the exterior
gravitational perturbations of an oscillating star have never been
solved numerically on such a grid.  Thus, the study of the pulsations
of rotating stars to date has been limited to the Newtonian
\cite{ipser-lind} and post-Newtonian \cite{cutler-lind} approximations.

The purpose of this paper is to explore the possibility of replacing
the outgoing-radiation boundary condition on the gravitational
potentials with a condition that is easier to implement numerically.
We derive conditions that are satisfied approximately by the outgoing
gravitational perturbations in the near zone of the gravitational
field.  These conditions can easily be imposed on the gravitational
fields in the well determined spacetime region just outside the star,
even in numerically determined models.  We propose that these
near-zone boundary conditions be imposed on the gravitational
potentials at the surface of the star as a substitute for the
outgoing-radiation boundary condition.  We test the accuracy of this
proposal by evaluating the non-radial modes of non-rotating stars,
which can also be determined exactly.  We find that the hydrodynamic
modes are determined using the near-zone boundary condition with
considerable precision: the real parts of the frequencies are
determined with an accuracy better than our knowledge of the exact
frequencies (about $0.01\%$) except in the most relativistic models
where it decreases to about $0.1\%$.  The imaginary
parts are determined with an accuracy of about $M/R$ for typical
neutron-star equations of state.

As part of this analysis we have derived a new representation of the
exact relativistic pulsation equations for non-rotating stars, which
is of considerable interest in its own right.  We reduce the equations
to a pair of second-order ordinary differential equations for two
scalar potentials: one fluid perturbation potential, and one
gravitational potential.  This representation of the equations is
exactly analogous to the Newtonian pulsation equations
\cite{ipser-lind}.  The analysis of the near-zone boundary conditions
described here is more straightforward using this new representation
of the pulsation equations.

Section~\ref{II} describes the new representation of the relativistic
pulsation equations in terms of two scalar potentials.
Section~\ref{III} presents the standard boundary conditions used to
determine the outgoing modes of relativistic stars in terms of the
two-potential formalism introduced here.  Section~\ref{IV} derives
approximate expressions for the gravitational potentials of the
outgoing modes that are valid in the near zone of the star.  These
expressions are used as the near-zone boundary conditions.  The
accuracy of these boundary conditions is tested by determining the
modes of a family of typical neutron-star like relativistic stellar
models.  Section~\ref{V} concludes with a discussion of the
feasibility of extending these near-zone techniques to rotating stars
in general relativity theory.  This paper also has two technical
appendices which give in detail the new two-potential forms of the
relativistic pulsation equations (Appendix A), and series solutions of these
equations that are useful for imposing the boundary conditions within
the star (Appendix B).

\section{Non-Radial Pulsations}
\label{II}

The pulsations of a relativistic stellar model are described by the
perturbations in the spacetime metric $\delta g_{ab}$, and the
associated perturbations in the stress-energy of the stellar matter
$\delta T^{ab}$.  The dynamical properties of these quantities are
determined by the perturbed Einstein equation,

\begin{equation}
\delta G^{ab} = 8\pi \delta T^{ab},\label{1}
\end{equation}

\noindent and the appropriate outgoing gravitational radiation boundary
condition at infinity.  Here we consider in detail the non-radial
pulsations of a spherical stellar model.  It is convenient to describe
the perturbations of the metric in such a spacetime using the
Regge-Wheeler gauge, in which the metric and its perturbations have the
following form,

\begin{eqnarray}\nonumber
&&(g_{ab}+\delta g_{ab})dx^a dx^b
=-e^\nu (1+H_0\,Y_{lm}\,e^{i\,\omega\,t})\,dt^2\\
&&\nonumber\quad+2\,i\,H_1\,Y_{lm}\,e^{i\,\omega\,t}\,dt\,dr
+ e^\lambda (1-H_0\,Y_{lm}\,e^{i\,\omega\,t})\,dr^2\\
&&\quad+ r^2 (1-K\,Y_{lm}\,e^{i\,\omega\,t})
\,(d\theta^2 +\sin^2\theta\, d\varphi^2).
\label{2}
\end{eqnarray}

\noindent The perturbation functions $H_0$, $H_1$, and $K$, and the
functions that describe the spacetime of the background equilibrium
star, $\nu$ and $\lambda$, depend only on the radial coordinate $r$. 
The $Y_{lm}$ are the standard spherical harmonics, and the constant
$\omega$ is the frequency of the mode.  The explicit expressions for the
components of the perturbed Einstein tensor $\delta G^{ab}$ that appear
in Eq.~(\ref{1}) have been given numerous times elsewhere
in terms of these fields \cite{ThorneCamp}, and will not be repeated
here.  We use standard geometrical units in which $G=c=1$. 

The perturbations in the stress-energy tensor $\delta T^{ab}$ are
assumed here to be those of a perfect fluid:

\begin{eqnarray}\nonumber
\delta T^{ab} = &&(\delta \rho + \delta p) u^a u^b 
+(\rho+p)(\delta u^a u^b + u^a \delta u^b)\\
&& + \delta p\, g^{ab}
- p g^{ac}g^{bd}\delta g_{cd},
\label{3}
\end{eqnarray}

\noindent where $\rho$, $p$, and $u^a$ denote the density, pressure, and
four velocity ($u^au_a = -1$) of the unperturbed equilbrium model; and
quantities preceeded by $\delta$ denote their Eulerian perturbations.
For simplicity here we restrict our attention to simple barotropic
perturbations, i.e., those where the density perturbation is
proportional to the pressure perturbation:

\begin{equation}
\delta \rho = {d\rho\over dp} \delta p={d\rho\over dp} \delta p(r)Y_{lm}
e^{i\omega t}
.\label{4}
\end{equation}
 
\noindent   The perturbed conservation law $\delta(\nabla_a
T^{ab})=0$ can be solved analytically to obtain a simple and useful
expression for the four-velocity perturbations.  This
solution is most conveniently expressed in terms of a scalar
perturbation function $\delta U$, defined as

\begin{equation}
\delta U(r)= {\delta p(r)\over \rho+p} 
+ \case1/2 H_0(r).\label{5}
\end{equation}

\noindent  Solving the spatial part of the perturbed conservation law 
gives an expression for $\delta u^a$ in terms of $\delta U$ and the
metric perturbations \cite{IpserLind}:

\begin{eqnarray}\nonumber
\delta u^a &&
= \case{i}{\omega} e^{\nu/2} e^{i\omega t} \nabla^a(\delta U\,Y_{lm})\\
&&+ \bigl(\case1/2 e^{\nu/2}H_0 \nabla^a t + e^{-\nu/2}H_1 \nabla^a r
\bigr)
Y_{lm} e^{i\omega t}.\label{6}
\end{eqnarray}

\noindent This expression allows us to describe a stellar oscillation
completely in terms of the four scalar functions $\delta U$, $H_0$,
$H_1$ and $K$. 

The perturbed Einstein equation  (\ref{1}) can be used to reduce further the
number of independent functions needed to determine the perturbed state
of the star.  The $\delta G{}^{\,r}{}_\theta$ and $\delta G{}^{\,t}{}_\theta$
components can be used to express $H_1$ and $H'_1$ in terms of the other
fields:

\begin{equation}
\omega e^{-\nu}H_1 = H'_0-K'+\nu'H_0,\label{7}
\end{equation}

\begin{eqnarray}\nonumber
\omega H'_1 =&&- \case1/2 \omega (\nu'-\lambda') H_1 
- \omega^2 e^\lambda(H_0+K)\\
&& -16\pi(\rho+p)e^{\nu+\lambda}\delta U.\label{7.1}
\end{eqnarray}

\noindent We use the notation ${}'$ to denote differentiation with
respect to $r$.  The $\delta G{}^{\,r}{}_r$ and the $\delta
G{}^{\,r}{}_t$ components together with Eqs.~(\ref{7}) and (\ref{7.1})
can similarly be used to express $K$ and $K'$ in terms of $H_0$,
$H'_0$, $\delta U$ and $\delta U'$:

\begin{equation}
K=\alpha_1 H'_0 + \alpha_2 H_0 + \alpha_3\delta U'+\alpha_4 \delta U,
\label{8}
\end{equation}

\begin{equation}
K'=\beta_1 H'_0 + \beta_2 H_0 + \beta_3 \delta U' + \beta_4 
\delta U,\label{9}
\end{equation}

\noindent where the functions $\alpha_i$ and $\beta_i$ depend on the
equilibrium structure of the star (and the frequency) and are given
explicitly in Appendix \ref{App.A}.  These expressions, Eqs.
(\ref{7})--(\ref{9}), can be used to eliminate $H_1$, $H'_1$, $K$ and
$K'$ from all of the remaining pulsation equations.  The equations
that determine the modes of relativistic stellar models can be reduced
therefore to a pair of second-order equations for the fields
$H_0$ and $\delta U$.  A second-order equation for $H_0$ is obtained
by using the derivative of Eq.~(\ref{7}) to eliminate $K''$ from the
$\delta G{}^{\,t}{}_t$ component of Einstein's equation.  A
second-order equation for $\delta U$ is obtained from the time
component of the perturbed stress-energy conservation law.  In each of
these equations all occurances of $H_1$ and $H'_1$ are eliminated with
Eqs.~(\ref{7}) and (\ref{7.1}), while $K$ and $K'$ are eliminated
using Eqs.~(\ref{8}) and (\ref{9}).  The resulting second-order
equations can be expressed in the form:

\begin{eqnarray}
\nonumber\delta U'' 
&&+ \left({2\over r}-{\nu'\over 2}{d\rho\over dp} 
+ \upsilon_1\right)\delta U'\\
&&\nonumber+\left[{\omega^2\over e^\nu}{d\rho\over dp} -{l(l+1)\over r^2}
+\upsilon_2\right]e^\lambda\delta U\\
&&=\upsilon_3 H'_0 
+ \left[{\omega^2\over 2e^\nu}{d\rho\over dp} + \upsilon_4\right]
e^\lambda H_0,\label{10}
\end{eqnarray}

\begin{eqnarray}
\nonumber H''_0&&+\left({2\over r} + \eta_1\right)H'_0\\
&&\nonumber +\left[{\omega^2\over e^\nu} - {l(l+1)\over r^2} 
+ 4\pi(\rho+p){d\rho\over dp}+\eta_2\right]e^\lambda H_0\\
&&=\eta_3\delta U' + \left[8\pi(\rho+p){d\rho\over dp}+\eta_4\right]
e^\lambda\delta U,\label{11}
\end{eqnarray}

\noindent where the functions $\upsilon_i$ and $\eta_i$ depend on the
structure of the equilibrium star (and the frequency) and are given in
Appendix \ref{App.A}. 

Equations (\ref{10}) and (\ref{11}) are the general equations for the
linear (even-parity) modes of general relativistic stellar models.  This
form of the equations has several advantages over earlier
representations of the fourth-order system needed to
describe these oscillations \cite{LindDet}, \cite{DetLind},
\cite{IpserPrice}.  Equations~(\ref{10}) and (\ref{11}) reduce in a
straightforward way, for example, to the Newtonian equations for the
non-radial modes in the appropriate weak-field slow-motion limit.
(Simply set $\upsilon_i=\eta_i=0$, $\nu'=-2p'/\rho$, $e^\nu=e^\lambda=1$
in Eqs.~[\ref{10}] and [\ref{11}], and $\omega^2=p=0$ in Eq.~[\ref{11}].)
The equation for the gravitational perturbation $H_0$ reduces to a
scalar wave equation in the region of spacetime far away from the
star, and in the static limit to Laplace's equation.  Thus it is
straightforward to determine the asymptotic behavior of the perturbed
fields from this representation of the equations.  The functions
$\upsilon_i$ and $\eta_i$ are smaller than the terms explicitly given
in Eqs.~(\ref{10}) and (\ref{11}) by a factor of order $M/R$, with $M$
the total mass and $R$ the total radius of the star.  Thus these
terms are small except in the interiors of extremely relativistic
models.  A simple and elegant form of the relativistic Cowling
approximation \cite{IpserLind} \cite{LindSplint} can be obtained by
setting $H_0=0$ in Eq.~(\ref{10}).  In this case the single fluid
potential $\delta U$ is determined by a single second-order wave
equation.  The solutions to this equation give fairly good
approximations of the hydrodynamic modes of relativistic stars
\cite{LindSplint}.  Similarly we presume that approximate expressions
for the even-parity $w$-modes (or `wave' modes \cite{Kokkotas}) could
be obtained by setting $\delta U=0$ in Eq.~(\ref{11}) and solving the
resulting homogeneous second-order equation for $H_0$.

\section{Boundary Conditions}
\label{III}

The physical solutions to Eqs.~(\ref{10}) and (\ref{11}) are
identified by imposing suitable boundary conditions at the center and
surface of the fluid interior of the star, and also in the asymptotic
wave zone of the gravitational field.  The boundary conditions
associated with the stellar fluid region are rather simple.  At the
center of the star the perturbations are simply required to be finite.
This condition eliminates one singular solution to the equations for
each of the functions $H_0$ and $\delta U$.  The non-singular
solutions can be approximated as power series near $r=0$,

\begin{equation}
\delta U = A r^l + {\cal O}(r^{l+2}),\label{12}
\end{equation}
\begin{equation}
H_0=Br^l  + {\cal O}(r^{l+2}),\label{13}
\end{equation}

\noindent where $A$ and $B$ are arbitrary constants.  

At the surface of the fluid region a boundary condition must also be
imposed that fixes the location of the perturbed surface of the star.
Those thermodynamic potentials that vanish continuously on the
surface of the equilibrium star (e.g. the pressure) must also vanish
on the moving surface of the perturbed star.  This condition is
equivalent to the requirement that the Lagrangian perturbation of
these potentials must vanish at the surface of the star \cite{Fried}.
The thermodynamic potential $h(p)$ defined by

\begin{equation}
h(p) = \int_0^p{dp'\over \rho(p') + p'}\label{14}
\end{equation}

\noindent is the ideal choice to implement this boundary condition 
\cite{hnote}.  The needed condition is traditionally written
$\Delta h =0$ on the surface of the star.  This condition can also be
re-expressed in terms of the Eulerian quantities as

\begin{equation}
\Delta h = \delta h +{ e^{\nu/2}\over i \omega }
\,\delta u^a\nabla_a h = 0,\label{15}
\end{equation}  

\noindent for the perturbations of a non-rotating star considered here.  
Since $\delta h = \delta p/(\rho+p)$, this expression can be
re-written in terms of the fields $\delta U$ and $H_0$:

\begin{eqnarray}\nonumber
\omega^2 &&e^{\lambda-\nu}\left(\delta U - \case{1}{2}H_0\right)
\\&&- \case{1}{2}\nu'\Bigl[\delta U'-(1-\beta_1)H_0'
-(\nu'-\beta_2)H_0\Bigr]=0.\label{16}
\end{eqnarray}  

\noindent In Appendix~\ref{App.B} power series expressions (valid
near the surface of the star) are presented for the fields $H_0$ and
$\delta U$ which satisfy this condition.  These series solutions are
useful for imposing the boundary condition numerically on a finite
grid having points near, but not on the actual surface of the star.

The boundary conditions associated with the fluid interior of the
star, Eqs.~(\ref{12}), (\ref{13}), and (\ref{16}), are sufficient to
fix a unique (up to overall scale) solution to Eqs.~(\ref{10}) and
(\ref{11}) for each value of the frequency $\omega$.  These conditions
do not however fix the frequency of the pulsation modes of these
stars.  Indeed, relativistic stars can oscillate at any frequency at
all if they are driven by incoming gravitational radiation of that
frequency.  The {\it outgoing} modes of relativistic stars are defined
by the additional condition that they contain no incoming
gravitational radiation.  This condition must be imposed as an
additional boundary condition on the gravitational potentials in the
asymptotic wave zone.

In the exterior of the star the gravitational potential $H_0$ is
determined by the vacuum ($\rho=p=\eta_3=\eta_4=0$) limit of
Eq.~(\ref{11}).  The behavior of the solutions in the asymptotic wave
zone are of particular interest to us.  Thus, we consider the limiting
form of this equation when $M/r \ll 1$, where $M$ is the total mass of
the star.  In this limit Eq.~(\ref{11}) reduces to the homogeneous
equation,

\begin{equation}
 H''_0+\left({2\over r} + \eta_1\right)H'_0
+\left[\omega^2 - {l(l+1)\over r^2} 
+\eta_2\right] H_0=0,\label{17}
\end{equation}

\noindent where $\eta_1$ and $\eta_2$ are given in this case by

\begin{equation}
\eta_1 = \case{4}{ r}
\Bigl\{(l-1)(l+2)\Delta\bigl[l(l+1)-2r^2\omega^2\bigr]-1\Bigr\}
,\label{18}
\end{equation}

\begin{equation}
\eta_2 = 16r^2\omega^4 \Delta,\label{19}
\end{equation}

\noindent and where $\Delta$ reduces to

\begin{equation}
{1\over\Delta}=\bigl[(l-1)(l+2)-2r^2\omega^2\bigr]
\bigl[l(l+1)-2r^2\omega^2\bigr]+4r^2\omega^2.
\label{20}
\end{equation}

\noindent Remarkably, the general solution to this equation 
has been found analytically \cite{ThorneIV}:

\begin{eqnarray}\nonumber
H_0 =&& 
-\biggl[r\omega {d\over d(r\omega)} 
+ 1 + \case{1}{2}l(l+1)-(r\omega)^2\biggl]\\
&&\qquad\qquad\times
{2i\omega^{l+1}\bigl[C j_l(r\omega) + D y_l(r\omega)\bigr]
\over l(l-1)(2l-1)!!}
,\label{21}
\end{eqnarray}

\noindent where $j_l$ and $y_l$ are spherical Bessel functions and
$C$ and $D$ are arbitrary constants.  In the asymptotic wave zone,
$r\omega \gg 1$, the Bessel functions have simple asymptotic forms:

\begin{eqnarray}
\nonumber\bigl[C &&j_l(r\omega) + D y_l(r\omega)\bigr]\,e^{i\omega t}\\
\nonumber&&={D+iC\over 2r \omega}\,e^{i\omega(t+r)-il\pi /2}+
{D-iC\over 2r \omega}\,e^{i\omega(t-r)+il\pi/2}\\
&&\quad+{\cal O}(r\omega)^{-2}.\label{22}
\end{eqnarray}

\noindent The condition that the solution contain no incoming radiation
reduces to the constraint $D=-iC$.  Thus the required boundary
condition on the gravitational potential $H_0$ is that it
approach (as $r\rightarrow \infty$) the asymptotic expression

\begin{equation}
H_0 = -{2Cr\omega^{l+2} e^{-i\omega r+ il\pi/2}\over l(l-1)(2l-1)!!}
[1+{\cal O}(r\omega)^{-1}],
\label{23}
\end{equation}

\noindent where $C$ is an arbitrary constant.

\begin{figure}
\centerline{\psfig{file=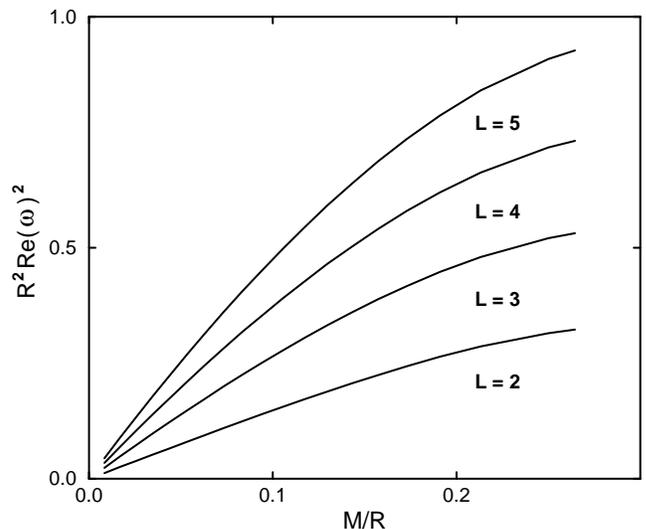,height=2.8in}}
\caption{Frequencies of the fundamental hydrodynamic modes
as a function of $M/R$ for typical relativisitic stellar models.}
\label{fig1}
\end{figure}

\section{Near-Zone Boundary Conditions}
\label{IV}

The frequencies of the hydrodynamic modes of interest to us here are
approximately related to the mass $M$ and radius $R$ of the star by
$R^2\omega^2 \approx l M/R$.  Figure~\ref{fig1} illustrates the
frequencies of these modes for ``typical'' relativistic stellar models
computed from the simplified neutron-star like equation of state
$p=\kappa \rho^2$, with $\kappa = 1.3346\times 10^5
\rm{cm}^5/\rm{g\,s}^2$.  (The maximum mass model for this equation of
state has $M=1.911M_\odot$, and the radius of the $1.4M_\odot$ model
is $R=14.45$km.)  The surfaces of these stars are approximately in the
near zone of the gravitational potential, $r^2\omega^2 < 1$, and also
in the zone (an extremely good approximation for the more Newtonian
models) where $M/r \ll 1$.  The expression for $H_0$ given in
Eq.~(\ref{21}) is valid in the entire exterior region of the star
where $M/r \ll 1$.  Thus, it is also valid in a portion of the near
zone for these modes.  In this region the expression for $H_0$ can be
simplified by employing the asymptotic expansions for the spherical
Bessel functions in the limit $r^2\omega^2
\ll 1$:

\begin{eqnarray}
H_0 &&={C\over r^{l+1}}\Bigl\{1 -i N_l(r\omega)^{2l+1} +
{\cal O}(r\omega)^{2}+{\cal O}(M/r)\Bigr\},\label{24}
\end{eqnarray}

\noindent where $N_l$ is given by

\begin{equation}
N_l=
 {(l+1)(l+2)\over l(l-1)(2l+1)[(2l-1)!!]^2}.\label{24.1}
\end{equation}

 The lowest-order version of the near-zone boundary condition sets
$H_0$ to the value given by Eq.~(\ref{24}):

\begin{equation}
H_0 ={C\over r^{l+1}}\Biggl[1 -i N_l(r\omega)^{2l+1}\Biggr].\label{25}
\end{equation}

\begin{figure} \centerline{\psfig{file=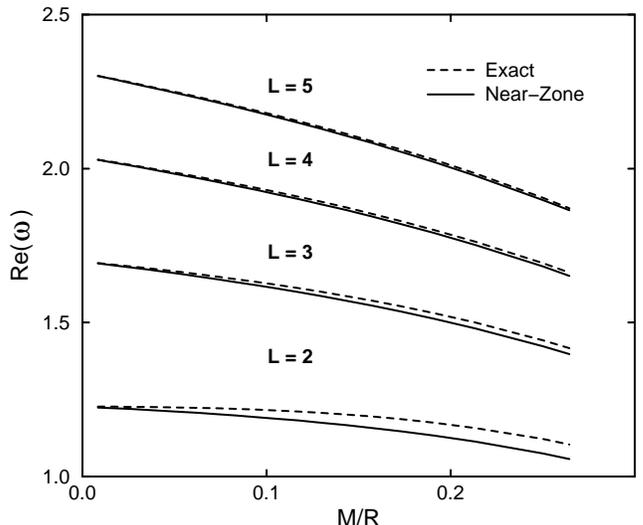,height=2.8in}} 
\caption{Real parts of the frequencies computed with the lowest-order 
near-zone boundary condition as given in Eq.~(\ref{25}).} 
\label{fig2}
\end{figure}

\begin{figure} \centerline{\psfig{file=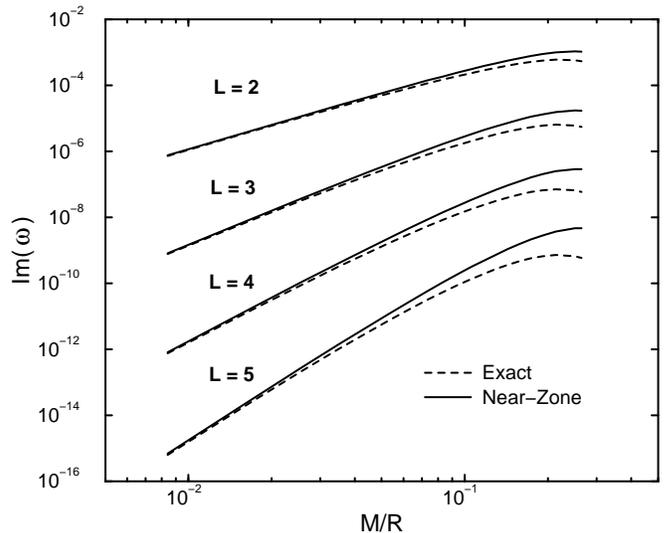,height=2.8in}} 
\caption{Imaginary parts of the frequencies computed with the 
lowest-order near-zone boundary condition as given in 
Eq.~(\ref{25}).} 
\label{fig3}
\end{figure}

\noindent The value of the derivative $H_0'$ is obtained by 
differentiation of Eq.~(\ref{25}).  The next question is, ``Where
should these conditions be imposed?''  Moving the boundary away from
the star improves the accuracy of the condition $M/r\ll 1$ used to
derive Eq.~(\ref{24}), but reduces the accuracy of the condition
$r^2\omega^2\ll 1$.  As Fig.~\ref{fig1} illustrates, the
$r^2\omega^2<1$ condition is just barely satisfied on the surfaces of
the most relativistic stellar models.  Thus, the most appropriate
place to impose the near-zone boundary condition, Eq.~(\ref{25}), is
directly on the surface of the star.  We have solved the relativistic
pulsation equations (\ref{10})--(\ref{11}) numerically using these
lowest-order near-zone boundary conditions for the stellar models
described above.  Figures~\ref{fig2} and \ref{fig3} illustrate the
real and imaginary parts of the frequencies determined in this way.
The frequencies in these figures are given in units of $\sqrt{M/R^3}$.
Also depicted in these figures are the frequencies obtained by solving
Eqs.~(\ref{10})--(\ref{11}) with the exact outgoing-radiation boundary
condition, Eq.~(\ref{23}), imposed far away from the star in the
wave zone.  These figures illustrate that the near-zone boundary
condition does reproduce in a qualitative way the frequencies of the
hydrodynamic modes of these stars.  For the more Newtonian models,
$M/R \ll 1$, the frequencies are determined quite accurately (as
expected).  For the more relativistic models, however, the agreement
is not as good: the real parts of the frequencies agree with the exact
to within about 4\%, while the imaginary parts only agree to within
about an order of magnitude.  This level of accuracy is, nevertheless,
somewhat better than that obtained with the standard
post-Newtonian approximation \cite{cutler-lind}.

  The expression for $H_0$ given in Eq.~(\ref{24}) includes only the
lowest-order terms in $(r\omega)^2$ and $M/r$.  By keeping additional
terms we could in principle construct a sequence of higher-order (and
one might expect more accurate) near-zone boundary conditions.  We
have found (after considerable numerical experimentation) that the
real part of $H_0$ is most sensitive to the $M/r$ terms in this
expansion, while the imaginary part is most sensitive to the
$(r\omega)^2$ terms.  This makes a certain amount of physical sense:
the real part of $H_0$ most strongly influences the real part of
$\omega$.  This hydrodynamic aspect of the mode is most sensitive to
the static ``Newtonian'' $M/r$ behavior of the potential.  The
imaginary part of $H_0$ determines, in effect, the imaginary part of
$\omega$.  This radiation-reaction aspect of the mode is most
sensitive to the dynamical $(r\omega)^2$ behavior of the potential.
Thus, we have constructed a boundary condition that keeps the $M/r$
terms to all orders for the real part of $H_0$, while keeping the
$(r\omega)^2$ terms to all orders for the imaginary part of $H_0$:

\begin{eqnarray}
\nonumber H_0 = &&C\Biggl\{
Q^2{}_l(x)\biggl[1 + {(l-8)(r\omega)^2\over 2l(2l-1)}\biggr]\\
\nonumber
&&-i K_l\biggl[r\omega {d\over d(r\omega)}+1+\frac{1}{2}l(l+1)
-(r\omega)^2\biggr]j_l(r\omega)\Biggr\},\\
\label{26}
\end{eqnarray}

\noindent where $C$ is an arbitrary constant, $x=r/M-1$, and $K_l$ is
given by

\begin{equation}
K_l = {2(l+2)!(M\omega)^{l+1}\over l(l-1)(2l-1)!!(2l+1)!!}.\label{27}
\end{equation}

\noindent The function $Q^2_l(x)$ is the associated Legendre function
of the second kind.  It is the exact solution to the static
($\omega=0$) limit of the $H_0$ equation, Eq.~(\ref{11}), that falls
off like $1/r^{l+1}$ for large values of $r$ \cite{ipser}.  Thus, the
real part of $H_0$ is given in Eq.~(\ref{26}) by an expression that is
exact to all orders in $M/r$ when $\omega=0$.  The lowest-order
$(r\omega)^2$ correction has also been added to this term.  The
imaginary part of $H_0$ in Eq.~(\ref{26}) is taken to be the exact
solution to the $H_0$ equation (\ref{11}) when $M=0$.  This is just
the spherical Bessel function solution that is given in Eq.~(\ref{21})
with the outgoing-radiation boundary condition.  This solution is
exact to all orders in $(r\omega)^2$ when $M=0$.  The constant $K_l$
was chosen so that the expression in Eq.~(\ref{26}) approaches
(\ref{25}) in the appropriate $M/r\ll 1$ and $(r\omega)^2\ll 1$ limit.

\begin{figure} 
\centerline{\psfig{file=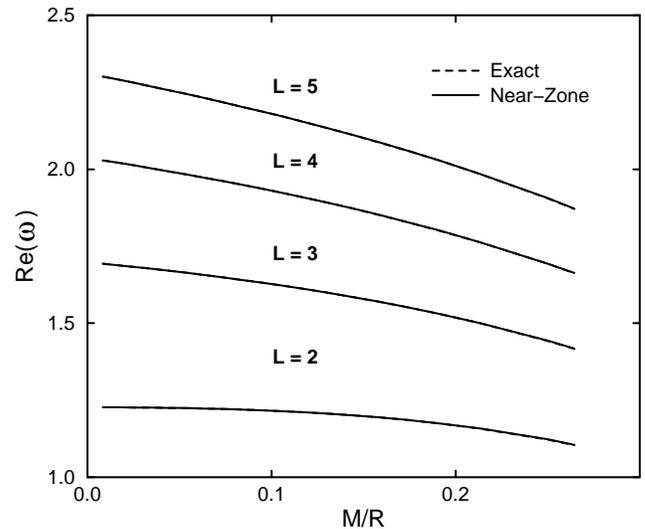,height=2.8in}} 
\caption{Real parts of the frequencies computed with the higher-order 
near-zone boundary condition as given in Eq.~(\ref{26}).} 
\label{fig4}
\end{figure}

\begin{figure} 
\centerline{\psfig{file=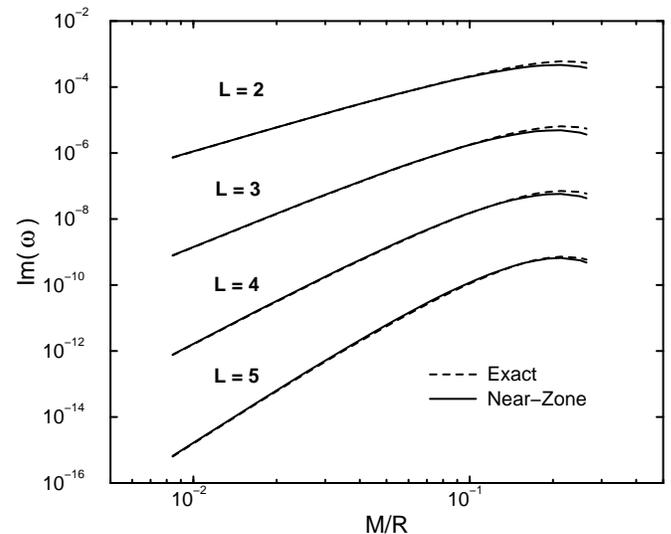,height=2.8in}} 
\caption{Imaginary parts of the frequencies computed with the 
higher-order near-zone boundary condition as given in 
Eq.~(\ref{26}).} 
\label{fig5}
\end{figure}

We have solved the relativistic pulsation equations
(\ref{10})--(\ref{11}) numerically using the higher-order near-zone
boundary condition given in Eq.~(\ref{26}) imposed at the surface of the
star $r=R$.  The derivative of $H_0$ is set to the derivative of
Eq.~(\ref{26}) at $r=R$.  Figures~\ref{fig4} and \ref{fig5} illustrate
the real and imaginary parts of the frequencies computed in this way.
The real parts of the frequencies computed with the higher-order
near-zone boundary conditions agree with those computed with the exact
outgoing-radiation boundary condition with impressive accuracy.  The
fractional error is less than about $0.01\%$, the numerical accuracy
with which we compute the exact frequencies, except in the most
relativistic models.  For the $l=2$ $f$-modes this error is about
$0.1\%$ in the maximum mass model, and about $0.02\%$ for the
$1.4M_\odot$ model.  The accuracy of the real parts of the frequencies
increases as $l$ increases.  The imaginary parts of the frequencies are
determined with accuracy of about $M/R$; considerably better than was
obtained with the lowest-order near-zone boundary condition.  The
maximum error is about 33\% for the $l=3$ mode of the most relativistic
model.  The accuracy for the modes of the $1.4M_\odot$ models is about
10\%.  This level of accuracy in the imaginary part of the frequency is
consistent with the fact that we have ignored terms of order $M/r$ in
Eq.~(\ref{26}) \cite{anote}.


To further illustrate the accuracy of the near-zone boundary conditions
we present in Table~\ref{TabI} the frequencies for the modes of the
$1.4M_\odot$ models computed with several different approximation
methods.  The Cowling approximation used here is obtained by setting
$H_0=0$ in Eq.~(\ref{10}).  The post-Newtonian frequencies presented
here are obtained from the $\gamma=2$ post-Newtonian frequencies
tabulated by Cutler and Lindblom \cite{cutler-lind} and adjusted to the
particular model considered here.  The near-zone boundary conditions
were used in both the lowest-order form of Eq.~(\ref{25}) and the
higher-order form of Eq.~(\ref{26}).

The detailed numerical results presented here were
obtained for the $\gamma=2$ polytropic equation of state described
above. We have also evaluated the modes of neutron star models based on
a softer $\gamma=5/3$ polytropic equation of state, and also on the more
realistic Bethe-Johnson  equation of state \cite{Bethe-Johnson}. We find
that the accuracy of the frequencies in these two additional cases is
comparable to (and in fact slightly better than) the $\gamma=2$
polytropic case discussed in detail here.

\section{Concluding Remarks}
\label{V} This paper presents a new formulation of the relativistic
pulsation equations in terms of two scalar potentials: one fluid and one
gravitational perturbation field.  These potentials satisfy a pair of
second-order ordinary differential equations that are analogous to the
Newtonian pulsation equations.  These equations are quite general: they
describe the complete set of even-parity modes of relativistic stars,
including the $w$-modes.  This formulation of the relativistic pulsation
equations has a significantly simpler structure than earlier
representations.

The primary purpose of this paper was to explore the possibility of
replacing the outgoing-radiation boundary condition imposed in the wave
zone with appropriate conditions imposed on the surface of the star.  We
have used the two-potential forms for the stellar pulsation equations to
derive the appropriate near-zone boundary conditions on the
gravitational potential.  We have shown that the outgoing hydrodynamic
modes can be determined with considerable precision using these
near-zone boundary conditions. Unfortunately this success is somewhat
vacuous: we already knew how to evaluate these modes exactly.  However
we believe that this proceedure should also allow us to evaluate the
modes of rapidly rotating fully relativistic stars with a precision
similar to that obtained here.  The near-zone boundary conditions that
we present here depend only weakly on the structure of the exterior
gravitational field of the star.  The boundary condition used to
determine the real part of the potential $H_0$ does depend on the exact
static solution to the exterior gravitational perturbation equations. 
But such static solutions have been found numerically in
rapidly-rotating strongly relativistic models \cite{nick}.  We believe
that these numerical solutions can be used to formulate a near-zone
boundary condition that will be accurate even for rapidly rotating
models.  The boundary condition on the imaginary part of $H_0$ does not
depend on the structure of the external gravitational field of the star
at all: it is just the flat-space perturbation representing an outgoing
wave.  Finally, the modes of primary interest in rotating stars have the
property that their frequencies $\omega$ decrease toward zero as the
angular velocity of the star increases.  For these modes the near-zone
approximation $r^2\omega^2\ll 1$ on which our boundary condition is
based will be satisfied more exactly as the angular velocity of the star
increases.

\acknowledgments

We thank C. Cutler, S. Detweiler, J. Friedman, B. Schutz, and K. Thorne
for helpful discussions concerning this work.  This research was
supported by NSF grants PHY-9796079 and PHY-9408910, NASA grants
NAG5-4093 and NAG5-3936, and the Institute for Fundamental Theory,
University of Florida.  L.L. thanks the Max Plank Institute f\"ur
Gravitationsphysik, Potsdam and the Institute for Fundamental Theory,
University of Florida for their hospitality during visits in
which portions of this research were completed.

\onecolumn
\widetext
\appendix

\section{The Pulsation Equations}
\label{App.A}

In Section \ref{II} are presented the general forms of the equations
that express $K$ and $K'$ in terms of $H_0$, $H'_0$, $\delta U$ and
$\delta U'$:

\begin{eqnarray}
K &=& \alpha_1 H'_0 + \alpha_2 H_0 + \alpha_3 \delta U' + \alpha_4 \delta U,
\label{a1}\\
K'&=& \beta_1 H'_0 + \beta_2 H_0 + \beta_3 \delta U' + \beta_4 \delta U.
\label{a2}
\end{eqnarray}

\noindent Here we present the complete expressions for the coefficients
$\alpha_i$ and $\beta_i$:

\begin{eqnarray}
\alpha_1 &=& -r \Delta \bigl\{4r^2\omega^2 e^{-\nu}
                         +r\nu'[16\pi r^2 (\rho+p)-l(l+1)]\bigr\},
\label{a3}\\
\nonumber\alpha_2 &=&\Delta \bigl\{ 
r(2-r\nu')[16\pi r^2(\rho+p)\nu'-l(l+1)\nu'+2r\omega^2e^{-\nu}]\\
&&-[l(l+1)e^\lambda - 2(1-r\nu')-8\pi r^2(\rho+p)e^\lambda]
[16\pi r^2 (\rho+p)-l(l+1)+2r^2\omega^2e^{-\nu}]\bigr\},
\label{a4}\\
\alpha_3 &=& -16\pi r^3 \Delta (\rho+p)(2-r\nu'),
\label{a5}\\
\alpha_4 &=& -16\pi r^2(\rho+p)e^\lambda \Delta [
16\pi r^2 (\rho+p)-l(l+1) +2r^2\omega^2 e^{-\nu}],
\label{a6}\\
\beta_1 &=& \Delta \bigl\{
[2r^2\omega^2 e^{-\nu}-(l-1)(l+2) ]
[16\pi r^2(\rho+p) - l(l+1)]e^\lambda
+2r^2\omega^2 e^{-\nu}(2-r\nu')\bigr\},
\label{a7}\\
\nonumber\beta_2&=&\Delta \bigl\{ r\omega^2 e^{-\nu}(2-r\nu')
[l(l+1)e^\lambda-2(1-r\nu')-8\pi r^2(\rho+p)e^\lambda]\\
&&\qquad-[(l-1)(l+2)-2r^2\omega^2 e^{-\nu}]
[16\pi r^2\nu'(\rho+p)-l(l+1)\nu' +2r\omega^2e^{-\nu}]e^\lambda\bigr\},
\label{a8}\\
\beta_3 &=& 16\pi r^2(\rho+p)\Delta e^\lambda 
[(l-1)(l+2) -2r^2\omega^2e^{-\nu}],
\label{a9}\\
\beta_4 &=& 16\pi r^3\omega^2(2-r\nu')(\rho+p)\Delta e^{\lambda-\nu},
\label{a10}
\end{eqnarray}

\noindent where $\Delta$ is defined by

\begin{equation}
\Delta^{-1} =
[(l-1)(l+2)-2r^2\omega^2e^{-\nu}][l(l+1)-2r^2\omega^2e^{-\nu}
-16\pi r^2(\rho+p)]e^\lambda
+ r^2\omega^2e^{-\nu}(2-r\nu')^2.
\label{a11}
\end{equation}

\noindent  This transformation becomes singular whenever $\Delta^{-1}$
vanishes.  We do not yet understand the general conditions under which
this singularity can occur (if any).  We have determined numerically,
however, that $\Delta>0$ for all $r$ in the hydrodynamic modes studied
here.

Also in Section \ref{II} were presented the general forms of the
second-order equations for $\delta U$ and $H_0$:

\begin{eqnarray}
\nonumber \delta U'' 
+ \left({2\over r}-{\nu'\over 2}{d\rho\over dp} 
+ \upsilon_1\right)\delta U'
&&+\left[{\omega^2\over e^\nu}{d\rho\over dp} -{l(l+1)\over r^2}
+\upsilon_2\right]e^\lambda\delta U
\\&&\qquad\qquad\qquad\qquad=\upsilon_3 H'_0 
+ \left[{\omega^2\over 2e^\nu}{d\rho\over dp} + \upsilon_4\right]
e^\lambda H_0,\label{a12}\\
\nonumber H''_0+\left({2\over r} + \eta_1\right)H'_0
+&&\left[{\omega^2\over e^\nu} - {l(l+1)\over r^2} 
+ 4\pi(\rho+p){d\rho\over dp}+\eta_2\right]e^\lambda H_0
\\&&\qquad\qquad\qquad
=\eta_3\delta U' + \left[8\pi(\rho+p){d\rho\over dp}+\eta_4\right]
e^\lambda\delta U.\label{a13}
\end{eqnarray}

\noindent Here we give the complete expressions for
the coefficients $\upsilon_i$ and $\eta_i$:

\begin{eqnarray}
\upsilon_1 &=&\nu'-\case1/2 \lambda' +
\biggl[{2\over r} -{\nu'\over 2}\biggl({d\rho\over dp}+1\biggr)\biggr]
\beta_3,
\label{a14}\\
\upsilon_2 &=&
16\pi(\rho+p)+e^{-\lambda}
\biggl[{2\over r} -{\nu'\over 2}\biggl({d\rho\over dp}+1\biggr)\biggr]
\beta_4,
\label{a15}\\
\upsilon_3 &=&
\biggl[{2\over r} -{\nu'\over 2}\biggl({d\rho\over dp} +1\biggr)\biggr]
(1-\beta_1),
\label{a16}\\
\upsilon_4 &=&-\case1/2\omega^2 e^{-\nu}+e^{-\lambda}
\biggl[{2\over r} -{\nu'\over 2}\biggl({d\rho\over dp} +1\biggr)\biggr]
(\nu'-\beta_2),
\label{a17}\\
\eta_1 &=&\case1/2 (\nu'-\lambda')
+ \case{2}{r}(2-r\nu')(\beta_1-1),
\label{a18}\\
\eta_2 &=& e^{-\lambda}
\bigl\{\nu'' + (\nu')^2 
+ \case{1}{2r}(2-r\nu')[4\beta_2-\nu'+\lambda']
\bigr\}+4\pi(\rho+p),\label{a19}\\
\eta_3 &=&-\case{2}{r}(2-r\nu')\beta_3,
\label{a20}\\
\eta_4 &=& -8\pi (\rho+p)
- \case{2}{r}(2-r\nu')e^{-\lambda}\beta_4.
\label{a21}
\end{eqnarray}

\twocolumn

\narrowtext

\section{Power Series Solutions}
\label{App.B}
  
It is helpful to have power series solutions to Eqs.~(\ref{10}) and
(\ref{11}) for the functions $\delta U$ and $H_0$ that satisfy the
appropriate boundary conditions near the center, $r=0$, and near the
surface, $r=R$, of the star.  Such series solutions are useful, for
example, in allowing the boundary conditions to be imposed numerically
at grid points near, but not on the actual boundaries.  Near $r=0$ the
situation is straightforward: the functions $\delta U$ and $H_0$ are
given by the series solutions

\begin{equation}
\delta U = A r^l + {\cal O}(r^{l+2}),\label{B1}
\end{equation}
\begin{equation}
H_0=Br^l  + {\cal O}(r^{l+2}),\label{B2}
\end{equation}

\noindent where $A$ and $B$ are arbitrary constants.

Near the surface of the star, however, the situation is more delicate. 
The thermodynamic derivative $d\rho/dp$ typically diverges at the
surface of the star.  Thus the second derivatives of $\delta U$ and
$H_0$ may be infinite at $r=R$.  This makes a straightforward power
series expansion impossible.  For definiteness we assume here that
in the neighborhood of the surface of the star the equation of state can
be represented as a polytrope

\begin{equation}
p = \kappa \rho^\gamma,\label{B3}
\end{equation}

\noindent where $\kappa$ and $\gamma$ are constants.  Near the surface
of such a star the equilibrium structure equations can be solved
as power series expansions.  From this we learn that

\begin{equation}
{d\rho\over dp} = {\rho\over \gamma p} 
= {R(R-2M)\over (\gamma-1)M(R-r)}[1+{\cal O}(R-r)],\label{B4}
\end{equation}

\begin{equation}
\rho=\biggl[ {\gamma -1\over \gamma\kappa}
{M(R-r)\over R(R-2M)}\biggr]^{1\over\gamma-1}
[1+{\cal O}(R-r)].\label{B5}
\end{equation}

\noindent Thus the derivative $d\rho/dp$ that appears as a coefficient
in Eq.~(\ref{10}) always diverges for polytropes.  The combination
$(\rho+p)d\rho/dp$ that appears in Eq.~(\ref{11}) diverges for stiff
equations of state ($\gamma>2$) but not for softer equations
of state:

\begin{equation}
(\rho+p){d\rho\over dp} = {1\over \kappa \gamma}
\biggl[ {\gamma -1\over \gamma\kappa}
{M(R-r)\over R(R-2M)}\biggr]^{2-\gamma\over\gamma-1}
[1+{\cal O}(R-r)].\label{B6}
\end{equation}

The boundary conditions determine the surface values of $\delta U'(R)$
and $H_0'(R)$.  The value of $H_0'(R)$ is determined by continuity at
$r=R$ from the outgoing solution in the exterior of the star.  The
value of $\delta U'(R)$ is determined by the boundary condition
Eq.~(\ref{16}).  These constants can be used therefore to give linear
expressions for the functions $H_0(r)$ and $\delta U(r)$ near the
surface:

\begin{equation}
H_0(r)= H_0(R)-(R-r)H'_0(R) + {\cal O}(R-r)^{1+\tau},
\label{B7}
\end{equation}

\begin{equation}
\delta U(r) = \delta U(R) - (R-r)\delta U'(R)  
+ {\cal O}(R-r)^{1+\tau},
\label{B8}
\end{equation}

\noindent where $0<\tau\leq 1$ is a positive constant that depends on
the equation of state.  For soft equations of state $1<\gamma<2$ the
singularities are relatively beneign, and the second derivatives $H''_0$
and $\delta U''$ are finite at $r=R$.  Thus, in this case series
expansions analogous to Eqs.~(\ref{B7}) and (\ref{B8}) can be obtained
for the derivatives:

\begin{equation}
H'_0(r)= H'_0(R) -(R-r)H''_0(R) + {\cal O}(R-r)^2,\label{B9}
\end{equation}

\begin{equation}
\delta U'(r)= \delta U'(R) -(R-r)\delta U''(R) + {\cal O}(R-r)^2,\label{B10}
\end{equation}

\noindent where the constants $H''_0(R)$ and $\delta U''(R)$ are
evaluated directly from the Eqs.~(\ref{10}) and (\ref{11}).  In the
stiff case, $\gamma > 2$, series expansions can also be found using
Eqs.~(\ref{B4})--(\ref{B6}) and the equations (\ref{10})--(\ref{11}). 
These expansions are not analytic at $r=R$ but can nevertheless be used
to approximate the derivatives $H'_0(r)$ and $\delta U'(r)$ near the
surface:

\begin{eqnarray}
\nonumber H'_0(r)&&= H'_0(R)
- \case{4\pi }{M}R^2\rho(r)[2\delta U(R)-H_0(R)]
\\&&\qquad\qquad\qquad\qquad\qquad
\times[1+{\cal O}(R-r)],
\label{B11}
\end{eqnarray}

\begin{eqnarray}
\nonumber &&\delta U'(r) = \delta U'(R)\\\nonumber &&
\quad-\case{1}{2}
\Bigl\{\case{4\pi}{M}R^2 [1-\beta_1(R)]\rho(r)[2\delta U(R)-H_0(R)]\\
&&\quad+\beta_3(r)\delta U'(R) +\beta_4(r)
\delta U(R)\Bigr\}[1+{\cal O}(R-r)].
\label{B12}
\end{eqnarray}

\narrowtext


\mediumtext
\begin{table} \caption{The exact frequencies of the
$f$-modes of $1.4M_\odot$ stellar models are compared with  those
computed using the near-zone boundary condition, and with those computed
using several other approximation methods.} 
 \begin{tabular}{ccccccc}
&$l$&Cowling&post-Newtonian&Lowest-Order&Higher-Order&Exact\\
&&&&Near-Zone&Near-Zone\\ \tableline
Re($\omega$)&2&1.4484&1.231&1.1672&1.2018&1.2016\\
Re($\omega$)&3&1.7469&1.619&1.5716&1.5863&1.5862\\
Re($\omega$)&4&1.9906&1.907&1.8652&1.8739&1.8738\\
Re($\omega$)&5&2.2031&2.147&2.1071&2.1130&2.1129\\
Im($\omega$)&2&--&--&$5.60\times10^{-4}$&$3.53\times10^{-4}$&$3.95\times10^{-4}$\\
Im($\omega$)&3&--&--&$7.29\times10^{-6}$&$3.56\times10^{-6}$&$3.96\times10^{-6}$\\
Im($\omega$)&4&--&--&$9.36\times10^{-8}$&$3.78\times10^{-8}$&$4.01\times10^{-8}$\\
Im($\omega$)&5&--&--&$1.09\times10^{-9}$&$3.70\times10^{-10}$
   &$3.63\times10^{-10}$\\
\end{tabular}
\label{TabI}
\end{table}


\begin{references}

\bibitem{Thorne} K. S. Thorne, Astrophys. J. {\bf 158}, 1 (1969).

\bibitem{LindDet} L. Lindblom and S. L. Detweiler, Astrophys. J. Suppl.
{\bf 53}, 73 (1983).

\bibitem{ChandraFerrari} S. Chandrasekhar and V. Ferrari, Proc. Roy. Soc.
(London) A {\bf 432}, 247 (1991).

\bibitem{Kokkotas} K. D. Kokkotas and B. F. Schutz, Mon. Not. Roy. Astro. Soc.
{\bf 255}, 119 (1992).

\bibitem{ipser-lind} J. R. Ipser and L. Lindblom, Astrophys. J. {\bf 355},
226 (1990).

\bibitem{cutler-lind} C. Cutler and L. Lindblom, Astrophys. J. {\bf 385}, 630
(1992).

\bibitem{ThorneCamp} K. S. Thorne and A. Campolattaro, Astrophys. J.
{\bf 149}, 591 (1967) and K. S. Thorne and A. Campolattaro,
Astrophys. J. {\bf 152}, 673 (1967).

\bibitem{IpserLind} J. R. Ipser and L. Lindblom, Astrophys. J. {\bf 389}, 392
(1992).

\bibitem{DetLind} S. Detweiler and L. Lindblom, Astrophys. J. {\bf 292}, 12
(1985).

\bibitem{IpserPrice} J. R. Ipser and R. H. Price, Phys. Rev. D {\bf 43}, 1768
(1991).

\bibitem{LindSplint} L. Lindblom and R. J. Splinter, Astrophys. J. {\bf 348},
198, (1990)

\bibitem{Fried} J. L. Friedman and B. F. Schutz, Astrophys. J. {\bf 200},
204 (1975).

\bibitem{hnote} The function $h(r)$ vanishes linearly at the surface
of any equilibrium stellar model.  This makes $h(r)$ the most
slowly vanishing themodynamic function at the stellar suface
having a finite gradient there.  Imposing the boundary condition
$\Delta h = 0$ at the surface of the star insures that the
Lagrangian changes in other more rapidly vanishing potentials
(such as the pressure) will also vanish there.

\bibitem{ThorneIV} K. S. Thorne, Astrophys. J. {\bf 158}, 997 (1969).

\bibitem{ipser} J. Ipser, Astrophys. J. {\bf 166}, 175 (1971).

\bibitem{nick} N. Stergioulas, Ph. D. Thesis, University of Wisconsin
at Milwaukee (1996); N. Stergioulas and J. L. Friedman, gr-qc/9705056.

\bibitem{anote} Unfortunately the $M/r$ dependence of the exact expression 
for $H_0$ is a very slowly converging series in $M/r$.  Adding the
first order $M/r$ correction to the imaginary part of Eq.~(\ref{26})
can be accomplished by multiplying that term by $1-lM/r$.  This
correction reduces the accuracy of the frequencies of the modes in the
most relativistic models.  Presumably a more accurate boundary
condition than Eq.~(\ref{26}) could be constructed by adding the
appropriate higher-order terms in $M/r$.

\bibitem{Bethe-Johnson} H. A. Bethe and M. Johnson, Nucl. Phys. A {\bf 230},
1 (1974).

\end{references}
\end{document}